\documentclass[10pt]{article} % For LaTeX2e
\usepackage[preprint]{tmlr}

\usepackage{amsmath,amsfonts,bm}

\def\eqref#1{equation~\ref{#1}}
\def\1{\bm{1}}

\DeclareMathAlphabet{\mathsfit}{\encodingdefault}{\sfdefault}{m}{sl}
\SetMathAlphabet{\mathsfit}{bold}{\encodingdefault}{\sfdefault}{bx}{n}

\usepackage{hyperref}
\usepackage{url}
\usepackage{multirow}
\usepackage{algorithm}
\usepackage{algorithmic}
\usepackage{tcolorbox}
\usepackage{subcaption}
\usepackage{booktabs}
\usepackage[absolute,overlay]{textpos}
\usepackage{eso-pic}

\AtBeginDocument{%
  }

\usepackage{xspace}
\usepackage{setspace}

\newcommand{\model}{R2Rank\xspace}

\newcommand{\sigone}{\ensuremath{^{\dagger}}}   % p < 0.05
\newcommand{\sigtwo}{\ensuremath{^{\ddagger}}}  % p < 0.01
\newcommand{\sigthree}{\ensuremath{^{\S}}}      % p < 0.001

\title{Reasoning to Rank: An End-to-End Solution for Exploiting Large Language Models for Recommendation}

\author{
\name
Kehan Zheng$^{1}$\thanks{Equal contribution.}\!,
Deyao Hong$^{1}$\footnotemark[1]\!,
Qian Li$^{2}$,
Jun Zhang$^{2}$,
Huan Yu$^{2}$,
Jie Jiang$^{2}$,
Hongning Wang$^{1}$\thanks{Corresponding author.}
\\\\
\addr
$^{1}$Tsinghua University,\;
$^{2}$Tencent Inc.
\\\\
\email
zhengkh24@mails.tsinghua.edu.cn, hongdy22@mails.tsinghua.edu.cn, hw-ai@tsinghua.edu.cn
}

\def\month{MM}  % Insert correct month for camera-ready version
\def\year{YYYY} % Insert correct year for camera-ready version
\def\openreview{\url{https://openreview.net/forum?id=XXXX}} % Insert correct link to OpenReview for camera-ready version

\begin{document}

\begin{textblock*}{\paperwidth}(2.4cm, 1.5cm)
\includegraphics[height=1.3cm]{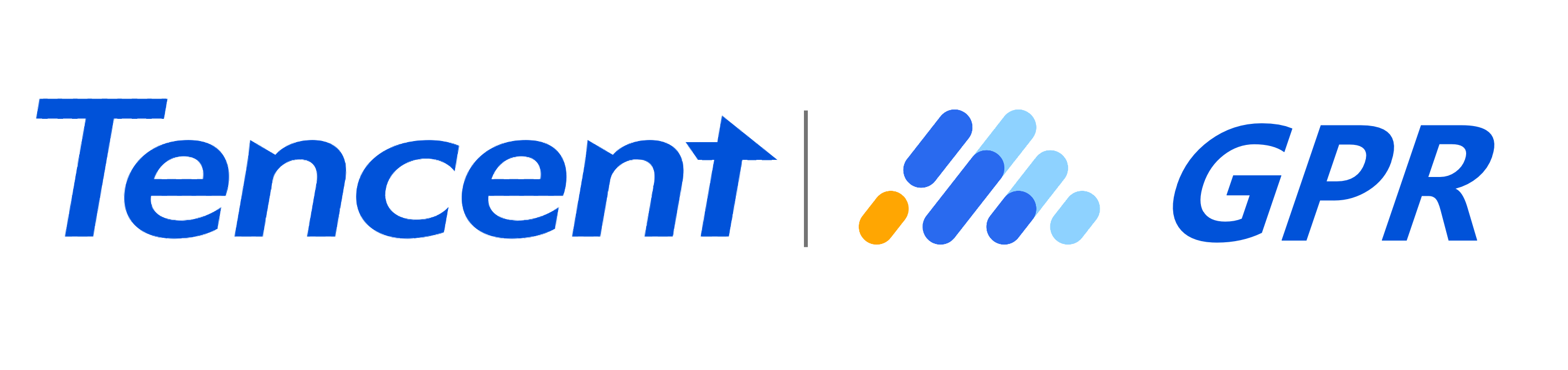} % 请确保目录下有 logo.png
\end{textblock*}

\maketitle

\begin{abstract}
Recommender systems are tasked to infer users' evolving preferences and rank items aligned with their intents, which calls for in-depth reasoning beyond pattern-based scoring. Recent efforts start to leverage large language models (LLMs) for recommendation, but how to effectively optimize the model for improved recommendation utility is still under explored. 
In this work, we propose Reasoning to Rank, an end-to-end training framework that internalizes recommendation utility optimization into the learning of step-by-step reasoning in LLMs. 
To avoid position bias in LLM reasoning and enable direct optimization of the reasoning process, our framework performs reasoning at the user-item level and employs reinforcement learning for end-to-end training of the LLM. 
%Given a user context and a candidate set, the LLM first produces a structured reasoning trace for each candidate item and then maps the reasoning into the final ranking of items. 
%This reasoning-driven recommendation procedure is optimized via reinforcement learning to ensure.
%To align reasoning-driven scores with ranking quality, we perform PPO-based learning-to-rank reinforcement learning with a Plackett–Luce surrogate,, enabling item-level credit assignment from listwise rewards (e.g., NDCG) and improving optimization stability, with an initial self-check SFT stage before RL. 
Experiments on three Amazon datasets and a large-scale industrial dataset showed consistent gains over strong conventional and LLM-based solutions. Extensive in-depth analyses validate the necessity of the key components in the proposed framework and shed lights on the future developments of this line of work.
\end{abstract}
\maketitle

\section{Introduction}
\label{sec:intro}

%Recommender systems have now become an indispensable outlet guiding users through an overwhelming number of choices in e-commerce, social media, and content platforms. At its core, recommendation is not merely a mechanical scoring or ranking task, but fundamentally an algorithmic decision-making process in which the system must infer users’ latent preferences based on their interaction history and produce recommendations aligned with their current needs and intents. As a result, high-quality recommendation inherently requires reasoning. Large language models (LLMs) are a good fit for this perspective because they can interpret rich text and perform multi-step reasoning, which supports preference understanding and intent-aware decision-making. This kind of step-by-step reasoning resembles System-2 thinking, making it possible to move beyond purely pattern-based recommendations.

Recommender systems have established themselves as an essential interface for users to navigate the overwhelming scale of modern digital ecosystems, such as e-commerce, social media, online education and many more \citep{RicciRokachShapira2022RSTechniques,AdomaviciusTuzhilin2005Survey}. However, conceptualizing recommendation merely as a mechanical ranking task underestimates the complexity of the challenge.
At its core, recommendation is essentially an inferential decision-making process, where a system must integrate fragmented evidence from a user's interaction history to reason about the user's evolving latent preferences and deduce their immediate intent 	\citep{RicciRokachShapira2022RSTechniques,KorenBellVolinsky2009MF}.
Consequently, achieving improved recommendation necessitates a paradigm shift from current shallow statistical pattern matching to deep logical reasoning \citep{Kahneman2011ThinkingFastSlow}.
Large Language Models (LLMs), with their remarkable capacity of pretrained world knowledge, semantic interpretation and multi-step reasoning, offer a unique opportunity to best realize this objective for recommendation~\citep{BrownEtAl2020GPT3,vaswani2017attention,WeiEtAl2022CoT,WuEtAl2024LLM4RecSurvey,collm2023,llmrec2023,p5tacl2022,tallrec2023,lin2024rellaretrievalenhancedlargelanguage,flowguided2024,reviewpref2025,you2025r2eclargerecommendermodels}.

%Recent work in the community has increasingly explored the use of LLMs for recommendation. Earlier studies mainly leveraged LLMs to construct content-based semantic features that complement conventional behavior signal driven recommenders. More recent approaches move toward end-to-end LLM recommenders, where the language model either produces item relevance scores as additional ranking signals or directly generates ranked lists. Most findings suggest the pretrained knowledge in LLMs exhibit strong potential in inferring user interest and their preferences on candidate items. 
%A central motivation behind this trend is to leverage, multi-step reasoning to better infer user interests from rich behavioral and textual evidence. 

In recognition of the potential, the research community has rapidly adopted LLMs to innovate recommendation methodologies ~\citep{WuEtAl2024LLM4RecSurvey,collm2023,llmrec2023,p5tacl2022,tallrec2023,lin2024rellaretrievalenhancedlargelanguage,flowguided2024,reviewpref2025,you2025r2eclargerecommendermodels}.
Earlier investigations mostly treated LLMs as auxiliary semantic encoders, utilizing their representations to augment behavior-based signals for recommendation ~\citep{collm2023,llmrec2023}.
Later, the focus has shifted toward building end-to-end LLM-based recommendation frameworks, where the model functions directly as the ranking agent -- either predicting numerical relevance scores or autoregressively generating ordered lists of candidates~\citep{p5tacl2022,tallrec2023, lin2024rellaretrievalenhancedlargelanguage}.
More recently, researchers identified that chain-of-thought reasoning in LLMs is also effective in discerning complex user interests and therefore improving recommendation quality~\citep{WeiEtAl2022CoT,flowguided2024,reviewpref2025, you2025r2eclargerecommendermodels}.
%Collectively, existing efforts demonstrate that the extensive world knowledge encapsulated within LLMs provides a robust basis for deciphering complex user interests and evaluating candidate relevance.

%for rank-style outputs, existing optimization often suffers from a structural mismatch between token-wise updates and sequence-level ranking signals, which leads to unreliable credit assignment across ranks and unstable list quality. Second, 
%Though promising, there are still several fundamental challenges left open for the LLM-based recommenders to be practically useful. First, LLMs are known victims of position bias, i.e., its output is sensitive to the presented order of recommendation candidates, which distorts the relationship between the model's reasoning and result ranking. Second, existing reasoning-augmented recommenders often rely on pre-defined prompt templates or weakly aligned reasoning supervision, where the generated reasoning traces can be superficial or misaligned with the recommendation objective, limiting how reliably reasoning translates into ranking gains.

Despite these advancements, fully exploiting LLMs' power and, especially, realizing the paradigm of reasoning for recommendation into practical effectiveness is still hindered by distinct structural obstacles ~\citep{liu2023lostmiddle,turpin2023unfaithfulcot}. 
First, LLMs are notorious victims of position bias, where the order of input disproportionately influences the model's output. This sensitivity breaks the permutation invariance required for robust item ranking, thereby distorting the assessment of true user preference in recommendation ~\citep{bito2025positionbiasllmrec}. 
Second, many approaches rely on carefully crafted prompts to guide LLMs to generate linguistically plausible reasoning traces but lack a principled mechanism that connects reasoning with the utility of the resulting recommendation. As a result, how to optimize LLM reasoning for improved recommendation quality is still underexplored.

To bridge the gap between LLMs' reasoning capabilities and the objective of recommendation, we propose Reasoning to Rank (\model), a framework that unifies Chain-of-Thought (CoT) reasoning with ranking optimization through an end-to-end reinforcement learning solution ~\citep{WeiEtAl2022CoT,SchulmanEtAl2017PPO}.
\model decouples the recommendation process into independent pointwise inferences: for each candidate item, the model first generates a rationale grounded in the user's interaction history and item information, and then projects this reasoning into a scalar relevance score -- a design that intrinsically mitigates position bias in LLMs.
To connect semantic reasoning with the optimization of ranking utility, we employ a Plackett-Luce probabilistic surrogate to map these scores into a differentiable ranking distribution~\citep{plackett1975analysis}.
This allows non-differentiable recommendation utility, usually measured by listwise rewards (e.g., NDCG), to be decomposed into item-level signals that back-propagate directly to the token-level reasoning content.
Furthermore, to address the scarcity of explicit reasoning in standard recommendation datasets, we introduce a cold-start Supervised Fine-Tuning (SFT) stage initialized via a self-reflective verification routine~\citep{MadaanEtAl2023SelfRefine}.
This holistic design ensures that the model's reasoning is not merely descriptive, but actively optimized to drive the recommendation performance in a permutation-invariant manner.

%%EXPERIMENT RESULTS%%
We conducted extensive experiments on three Amazon datasets and a large-scale industrial advertising dataset to compare \model with conventional recommenders \citep{kang2018selfattentivesequentialrecommendation, sun2019bert4recsequentialrecommendationbidirectional, he2020lightgcnsimplifyingpoweringgraph} and competitive LLM-based baselines \citep{Guo_2025,Zhang_2023, tallrec2023, you2025r2eclargerecommendermodels}. The results show that \model achieves the best or near-best performance across datasets on NDCG metrics,  validating the effectiveness of optimizing LLM reasoning for listwise recommendation utility. We further perform ablation studies to verify the contributions of key components in \model, and conduct in-depth analyses on cross-domain generalization, item cold-start robustness, and sensitivity to user history length, demonstrating that \model remains effective and stable across diverse settings.

Our contributions are summarized as follows. 
First, we propose a new end-to-end recommendation framework that internalizes recommendation utility optimization into the learning of step-by-step reasoning in LLMs. 
Second, we introduce a cold-start supervised fine-tuning strategy with a self-reflective routine to teach the LLM a stable and effective reasoning mode for inferring user interests. 
Third, we develop a reinforcement learning post-training method based on a Plackett–Luce differentiable surrogate, enabling rank-level credit assignment from listwise rewards (e.g., NDCG) to the detailed next token generation in the LLM. 
Fourth, extensive experiments on public benchmarks and large-scale industrial data demonstrate consistent gains over competitive baselines. Putting together, our efforts in this work shed lights on the future development of this line of recommendation solutions.

\section{Related Work}

We organize our discussion of related work into two parts: the recent efforts that leverage LLMs for recommendation and the developments of LLM post-training. 

\noindent\textbf{$\bullet$ Large Language Models in Recommender Systems} LLMs have recently been introduced into recommender systems as a flexible substrate for semantic understanding, instruction following, and multi-step reasoning, enabling recommendation to be formulated and solved through natural-language interfaces~\citep{WuEtAl2024LLM4RecSurvey}.
Existing studies can be summarized into several paradigms.
Some treat LLMs as prompt-based recommenders that directly select or rank candidates from a user context, but pure zero-shot prompting is often brittle and tends to underperform strong conventional rankers, especially under long contexts and larger candidate sets~\citep{liu2023chatgptgoodrecommenderpreliminary,liu2023lostmiddle}.
Others use LLMs offline as semantic feature enhancers to summarize histories, enrich item semantics, or synthesize preference descriptions that are then consumed by conventional ranking models, injecting language understanding while avoiding online serving latency~\citep{collm2023,llmrec2023,Liu_2024}.

More recent work moves toward end-to-end LLM recommenders by training the LLM itself to produce recommendation decisions, either via generative identifier/sequence prediction or via scoring-based ranking over candidates~\citep{p5tacl2022,tallrec2023,lin2024rellaretrievalenhancedlargelanguage,lin2025orderagnosticidentifierlargelanguage}.
In parallel, an emerging line explicitly leverages reasoning for recommendation, where chain-of-thought deliberation, review-grounded preference inference, or process-aware optimization is used to better connect textual rationales with ranking behavior~\citep{WeiEtAl2022CoT,reviewpref2025,flowguided2024,you2025r2eclargerecommendermodels}.
While these results suggest that reasoning can strengthen preference inference, broadly applicable objectives that consistently translate reasoning improvements into listwise ranking gains remain underexplored.

\noindent\textbf{$\bullet$ Post-training for Large Language Models}
Post-training has become a standard recipe for adapting pretrained LLMs, typically consisting of supervised fine-tuning (SFT) and preference-based optimization.
Representative approaches include policy-gradient RL methods (e.g., PPO) and direct preference optimization methods (e.g., DPO), as well as more recent group-based RL variants (e.g., GRPO)~\citep{ouyang2022traininglanguagemodelsfollow,SchulmanEtAl2017PPO,rafailov2024directpreferenceoptimizationlanguage,Guo_2025}.
These methods primarily align LLMs at the response level, optimizing how a single generated output matches a preference signal. 
In contrast, our post-training follows a learning-to-rank perspective: the optimization target is inherently listwise, aiming to align model behavior with ranking quality rather than solely improving standalone responses.

\begin{figure*}[!t]
  \centering
  \includegraphics[width=\linewidth]{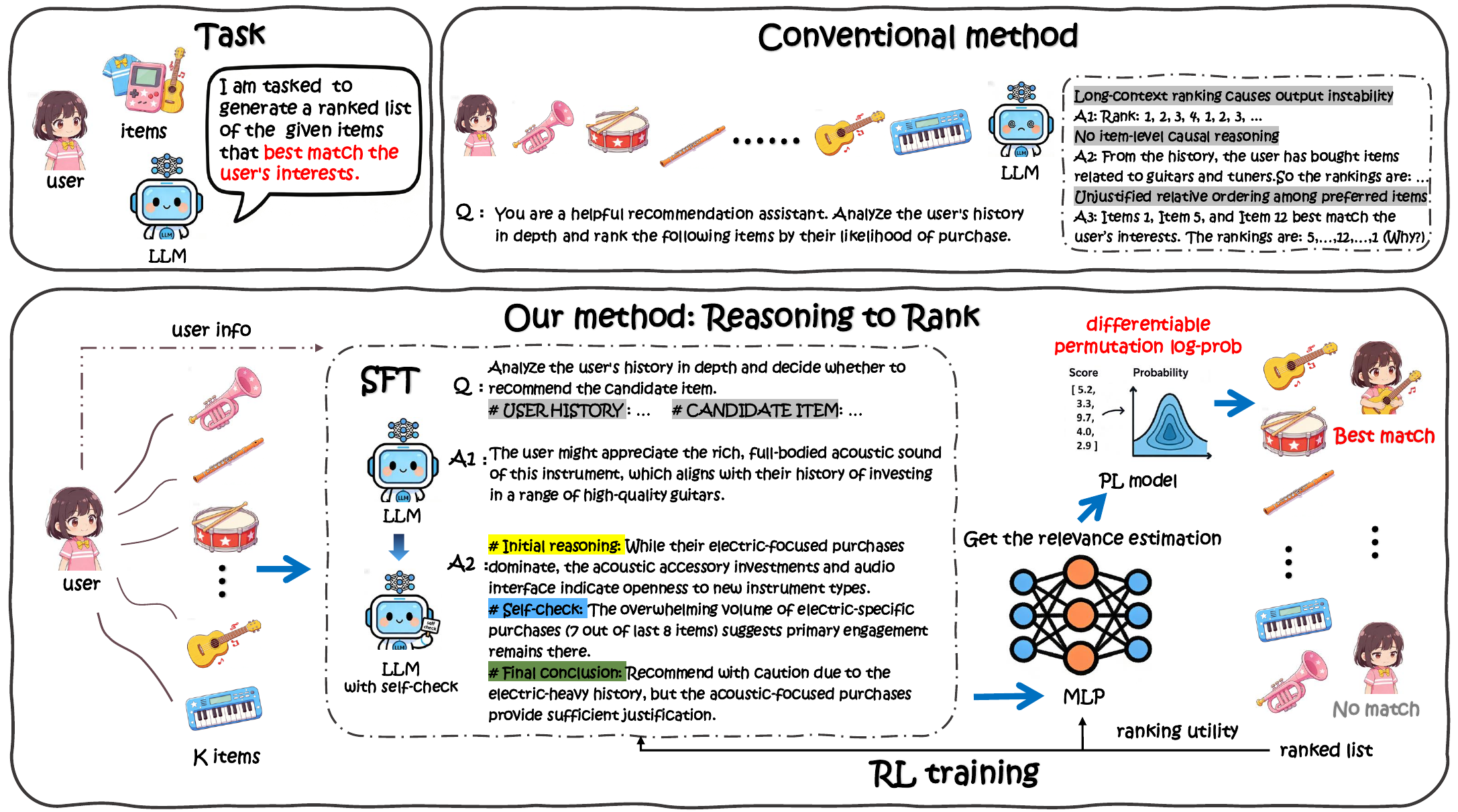}
  \vspace{2pt}
  \caption{Overview of the proposed Reasoning to Rank framework. The LLM performs item-level reasoning for each candidate item and aggregates them to produce the final ranking. The reasoning process is trained end-to-end via reinforcement learning for directly optimizing the final recommendation utility.}
  \label{fig:framework}
\end{figure*}

\section{Methodology}

In this section, we present our Reasoning to Rank (\model) framework in detail. 
We begin by specifying the reasoning-based recommendation setting and the notations used throughout this paper. 
Then, we provide an architectural overview, illustrating how pointwise item-level reasoning is synthesized into final recommendation rankings.
Finally, we provide the complete technical details about our developed end-to-end reinforcement learning solution and the self-reflective supervised fine-tuning construct to cold-start the model's recommendation-focused reasoning output.

%We present the Reasoning to Rank (\model) framework, beginning with the problem formulation and architectural overview. We then detail the core technical contributions: our end-to-end reinforcement learning optimization and the self-reflective fine-tuning strategy for cold-starting recommendation capabilities.

\subsection{Preliminaries}
\label{sec:prelim}

For each user $u\in\mathcal{U}$, we represent the user's observed interaction history as a chronologically ordered sequence of items,
$H_u = \langle v_1, v_2, \ldots, v_L\rangle$ with $v_\ell\in\mathcal{I}$, where each interacted item $v_\ell$ is depicted with descriptive signals, such as timestamp, textual item description and other forms of meta data.
When available, we denote the basic information (e.g., profile attributes) of a user as $b_u$.
The recommendation context for user $u$ is then defined as $c_u = (b_u, H_u)$.
Given a set of recommendation candidates $X_u=\{x_1,\ldots,x_K\}\subset\mathcal{I}$ for user $u$, a model (denoted as $\pi_\theta$) outputs a ranking
$\tau_u$ over $X_u$ to maximize the recommendation utility $\rho(\tau_u)$. 

In this paper, we focus on Normalized Discounted Cumulative Gain (NDCG) \citep{jarvelin2002cumulated} as the utility metric $\rho(\cdot)$, but the training framework can be easily adopted to all other ranking-based metrics. Specifically, denote the user-provided relevance feedback on $\tau_u$ as $R_u=\{r_1,\ldots,r_K\}$, where $r_k$ is a graded relevance label for item $x_k$ (e.g., 0/1 for binary feedback), NDCG is computed as, 
\begin{equation*}
    \text{NDCG}(\tau_u) = \frac{\text{DCG}(\tau_u)}{\text{DCG}(\tau^*_u)}
\end{equation*}
where $\text{DCG}(\tau)=\sum^K_{k=1}\frac{2^{r_{\tau(k)}}-1}{\log_2(k+1)}$ and $\tau^*_u$ stands for the ideal ranking of $X_u$, i.e., items ranked according to $R_u$.
%In existing LLM-based recommenders, the ranking over $X_u$ is typically produced in one of two ways: (i) listwise generation, where the model directly generates an ordered list in text; or (ii) candidate scoring, where the model produces a relevance score for candidates conditioned on the same context, and the final order is induced by sorting these scores.

\subsection{Framework Overview}

Simply feeding $c_u$ and $X_u$ into an LLM for a ranked list of recommendations is clearly not ideal. As previously discussed, item position in $X_u$ can inadvertently affect the LLM's output, though this is clearly irrelevant to the utility of recommendation candidates. Moreover, the growing size of $c_u$ and $X_u$ not only increases the difficulty of reasoning about the user's true preference, but also hinders the practicality of LLMs for recommendation, due to the prohibitive cost of computing full attention \citep{vaswani2017attention}.

As a result, our \model{} framework explicitly separates item-level user preference reasoning from ranking aggregation. 
As illustrated in Figure \ref{fig:framework}, for each item $x_i \in X_u$, the LLM is first tasked to produce structured reasoning $y_i$ conditioned on $c_u$, 
\begin{align}
y_i \sim \pi_\theta(\cdot \mid c_u, x_i), \qquad i=1,\ldots,K.
\end{align}
Crucially, the generation of $y_i$ is trained to encourage the model to extract rationales from the interaction history $c_u$ and connect them to the descriptive information about the candidate item to decide the item's relevance, rather than relying on item-agonistic or shallow shortcuts.

Then we map the reasoning content $y_k$ to a relevance score $s_k$ to form the final ranking $\tau_u$, 
\begin{equation}\label{eq:sort}
\tau_u \;=\; \mathrm{argsort}\big(\{s_k=f_\phi(y_k)\}_{k=1}^K\big),  
\end{equation}
where $f_\phi(\cdot)$, parameterized by $\phi$, is a scoring head directly connected to the last layer of $\pi_\theta$.

This staged decision making yields two practical benefits aligned with the challenges discussed in Section~\ref{sec:intro}. 
It avoids entangling reasoning about all candidates into a single decoding trajectory, which reduces the model's sensitivity to the order of candidates in the model input and prevents the ranking decision from drifting with the ordering or formatting of $X_u$. 
It also alleviates the long-context reasoning bottleneck by focusing deliberation on one candidate at a time under the given recommendation context $c_u$, encouraging the model to explicitly trace how signals in $c_u$ support (or contradict) recommending $x_i$. 

In the next section, we describe the end-to-end training solution that optimizes the generation of reasoning content for improved recommendation utility.

%Step 2 aggregates these judgments into a listwise ranking through scalar relevance scores. To make the free-form judgment operational, we attach a lightweight MLP scoring head $f_\phi(\cdot)$ and compute a scalar score from the representation of $y_i$. Concretely, we extract the hidden state of the last token of the final answer segment in $y_i$ as its representation and compute
%\begin{align}
%s_i = f_\phi\!\big(\mathrm{LastHidden}(y_i)\big).
%\end{align}
%The final ranking is induced by sorting these scores:

\subsection{Reinforcement Learning under Plackett-Luce Surrogate}

Once $\tau_u$ is presented to the user, user feedback $R_u$ can be collected to measure the utility of $\tau_u$. In end-to-end training, the optimization of the reasoning process should be driven by the recommendation utility, which requires the gradient to be back-propagated from $\rho(\tau_u)$ all the way to $\theta$ and $\phi$, via $\tau_u$, $\{s_k\}^K_{k=1}$ and $\{y_k\}^K_{k=1}$. 
In particular, the forward pass in \model{} is summarized below,
\begin{align}
(c_u,X_u)
\xrightarrow{\ \pi_\theta\ }
\{y_i\}_{i=1}^K
%\xrightarrow{\ \text{extract}\ }
%\{h_i\}_{i=1}^K
\xrightarrow{\ f_\phi\ }
\{s_i\}_{i=1}^K
\xrightarrow{\ \mathrm{argsort}(\mathbf{s})\ }
\tau_u
\xrightarrow{\ R_u \ }
\rho({\tau_u}). \label{eq:r2r_traj}
\end{align}
%Here, $\pi_\theta$ generates a structured preference judgment $y_i$ for each candidate $x_i$ under the same context $c_u$; $h_i$ is extracted from $y_i$ as the hidden state of the last token of the final answer segment; $f_\phi$ maps $h_i$ to the scalar relevance score $s_i$; $\tau$ is a permutation sampled from the PL distribution parameterized by $\mathbf{s}$; and $R$ is the resulting listwise reward. At test time, we drop the sampling step and output $\mathrm{argsort}(\mathbf{s})$.

Two technical barriers prevent direct gradient-based optimization of this pipeline. 
First, the ranking operation in Eq. \eqref{eq:sort} is non-differentiable, such that direct gradient back-propagation breaks at this step. 
Second, most metrics of recommendation utility, e.g., NDCG or hit rate, are discrete and thus non-differentiable either. 

In this work, we appeal to the Plackett-Luce model \citep{plackett1975analysis} to map the relevance scores to a distribution of rankings, which smooths the ranking operation in Eq. \eqref{eq:sort} and reinstates gradient back-propagation. 
On top of this design, we adopt policy-based reinforcement learning to optimize the final utility metric. According to the policy gradient theorem \citep{sutton1999policy}, one should follow the gradient direction weighted by the expected return under the current policy for policy improvement. Under the Plackett-Luce model, in \model{}, the expected return can be easily obtained via Monte Carlo sampling. In the following, we provide the technical details of these two treatments accordingly. 

%Our post-training objective is to directly optimize listwise ranking quality while jointly updating the LLM reasoning policy and the lightweight scoring head. The core difficulty in doing so is not the evaluation metric itself, but the inference-time ranking operator: given the predicted score vector $\mathbf{s}=(s_1,\ldots,s_K)$, the output ranking is obtained by a hard $\mathrm{argsort}(\mathbf{s})$, i.e., a composition of $\arg\max$ decisions that is non-differentiable with respect to $\mathbf{s}$. As a result, na\"ively differentiating through the inference pipeline provides no learning signal to the scores. To bypass this obstacle in a principled and learning-to-rank-compatible manner, we introduce a Plackett--Luce (PL) sampling surrogate during training. Importantly, PL sampling is used only for training; inference remains unchanged and deterministically returns $\mathrm{argsort}(\mathbf{s})$.

\begin{algorithm}[t]
\caption{RL training for \model{}}
\label{alg:r2r_rl}
\begin{algorithmic}[1]
\REQUIRE Minibatch $\mathcal{B}$; policy $\pi_{\theta}$; scoring head $f_\phi$; clipping ratio $\epsilon$.
\FOR{each instance $(u,c_u,X_u=\{x_1,\ldots,x_K\}) \in \mathcal{B}$}
    \FOR{$k=1$ to $K$}
        \STATE Sample $y_k \sim \pi_{\theta}(\cdot \mid c_u,x_k)$
        \STATE Compute $s_k \leftarrow f_\phi(y_k)$
    \ENDFOR
    \STATE Sample $\tau \sim P(\cdot|\boldsymbol{s})$ using Eq.~\eqref{eq:pl_dist}
    \STATE Evaluate $\rho(\tau) \leftarrow \mathrm{NDCG}(\tau)$
    \STATE Accumulate PPO objective $\mathcal{J}(\theta)$ using Eq.~\eqref{eq:ppo_obj}
    \STATE Accumulate REINFORCE objective $\mathcal{J}(\phi) \leftarrow  \rho(\tau) \log P(\tau|\boldsymbol{s})$
\ENDFOR
\STATE Update $\theta$ by ascending $\mathcal{J}(\theta)$
\STATE Update $\phi$ by ascending $\mathcal{J}(\phi)$
\end{algorithmic}
\end{algorithm}

\noindent\textbf{$\bullet$ Plackett-Luce Surrogate.}
Given $\boldsymbol{s}=\{s_k\}_{k=1}^K$, the Plackett-Luce (PL) model imposes a distribution of rankings over the $K$ items as, 
\begin{align}
P(\tau|\boldsymbol{s}) = \prod_{k=1}^{K}
\frac{\exp\left({s_{\tau(k)}}\right)}{\sum_{j=k}^{K}\exp\left({s_{\tau(j)}}\right)}. \label{eq:pl_dist}
\end{align}
At training time, we obtain $\tau_u$ by sampling from $P(\tau|\boldsymbol{s})$. 
Intuitively, sampling from Eq. \eqref{eq:pl_dist} turns the hard ranking operation into a stochastic policy over permutations, so that changes in $\mathbf{s}$ smoothly adjust the resulting rankings, enabling the gradient to pass through.

\noindent\textbf{$\bullet$ End-to-End Reinforcement Learning.}
%Because the reward depends on the sampled permutation, we optimize the scoring head through the log-derivative trick. Let $b$ be a baseline and $A=R-b$ be the advantage. 
The policy gradient of language model's parameter $\theta$ and scoring head's parameter $\phi$ can be computed in a similar manner. First, the expected ranking utility with respect to $\phi$ is computed as,
\begin{align}
\nabla_\phi \ \mathbb{E}_{\tau_u\sim P(\cdot|\boldsymbol{s})}[\rho(\tau_u)]
=
\mathbb{E}_{\tau\sim P(\cdot|\boldsymbol{s})}
\Big[
\rho(\tau_u) \,\nabla_\phi \log P(\tau_u|\boldsymbol{u})
\Big], \label{eq:pl_reinforce}
\end{align}
which can directly fed into the popularly used REINFORCE algorithm \citep{sutton1999policy} for the optimization of $\phi$.

The situation is a bit different for policy-based optimization of language model $\pi_\theta$. 
Denote the token sequence generated for candidate $x_k$ as $y_k=(o_{k,1},\ldots,o_{k,T_i})$. Using the language of Markov Decision Process, the state at position $t$ is the tuple $(c_u, x_k, o_{k,<t})$ and the action of policy $\pi_\theta$ is the choice of token $o_{k,t}$. However, the corresponding reward is delayed until the final ranking is assessed by the user. Simply applying REINFORCE on the optimization of $\theta$ inevitably suffers from high variance, and we appeal to the Proximal Policy Optimization (PPO) algorithm \citep{SchulmanEtAl2017PPO} as the remedy.

% \newpage% 让后面的内容从下一栏的顶部开始
% \noindent
\begin{figure}[t]
\begin{tcolorbox}[colback=gray!5,colframe=gray!60,title={Self-check Prompt Template },fonttitle=\bfseries]
\small
You are a recommendation assistant. Analyze the user's history in depth and decide whether to recommend the candidate item.\\
\textbf{Requirements:} Generate a response with the following three parts in order:\\
\#Initial reasoning: thoroughly analyze relevant evidence from \textsc{User Context} and \textsc{Candidate Item}.\\
\#Self-check: identify mistakes, missing evidence, uncertainties, or over-weighted signals in the initial reasoning.\\
\#Final conclusion and suggestion: produce a concise final rationale and decision.\\
Finally output: \texttt{<answer>Recommend</answer>} or \texttt{<answer>Not Recommend</answer>}.\\[2mm]
\textbf{User Context:} \{user\_profile\}, \{user\_history\}\\
\textbf{Candidate Item:} \{item\_text\}
\end{tcolorbox}
\caption{The self-reflective prompt template adopted in \model.}
\label{fig:selfcheck_template}
\end{figure}

Specifically, PPO performs policy gradient update in a trust region specified by the following objective function,
\begin{align}
\mathcal{J}_{\mathrm{PPO}}(\theta)
=
\sum_{k=1}^{K}\sum_{t=1}^{|y_k|}
\min\Big(
\gamma_{k,t}(\theta)\rho(\tau),\;
\mathrm{clip}(\gamma_{k,t}(\theta),1-\epsilon,1+\epsilon)\rho(\tau)
\Big), \label{eq:ppo_obj}
\end{align}
where $\epsilon$ controls the diameter of the trust region and $\gamma_{k,t}(\theta)$ is the token-level probability ratio,
\begin{align*}
\gamma_{k,t}(\theta)
=
\frac{\pi_\theta(o_{k,t}\mid c_u,x_i,o_{k,<t})}
{\pi_{\theta_{\mathrm{old}}}(o_{k,t}\mid c_u,x_i,o_{k,<t})}, \label{eq:ppo_ratio}
\end{align*}
in which $\theta_{\mathrm{old}}$ denotes the policy where the reasoning content $\{y_k\}^K_{k=1}$ is sampled. 

%Intuitively, PPO encourages reasoning trajectories whose induced score vectors yield sampled permutations with higher NDCG, while preventing overly large policy updates that harm training stability.

%\textbf{Overall objective.} Combining the two components, our post-training optimizes a single listwise signal through two coupled updates: (i) a PL-based REINFORCE update for the scoring head that shapes the permutation distribution induced by $\mathbf{s}$, and (ii) a PPO update for the LLM that improves the candidate-wise judgments that produce these scores. Concretely, for each instance we maximize
%\begin{align}
%\mathcal{J}(\theta,\phi) = \mathcal{J}_{\mathrm{PPO}}(\theta) + \lambda \, A \log P(\tau;\boldsymbol{\alpha}), \label{eq:joint_obj}
%\end{align}
%where $\lambda$ balances the scoring-head update against the PPO objective.

%\textbf{Training algorithm.}
Algorithm~\ref{alg:r2r_rl} summarizes mini-batch update for $\phi$ and $\theta$. In each user, we first sample the reasoning trajectory $y_k$ for every recommendation candidate $x_k$ following the current policy $\pi_\theta$ and compute the corresponding relevance score $\mathbf{s}=\{s_k\}^K_{k=1}$. Then we sample a ranking $\tau$ via from the PL surrogate and obtain the listwise reward $\rho(\tau)$. Finally, we update $\theta$ and $\phi$ via their gradients computed under REINFORCE and PPO accordingly.

\subsection{Self-reflective SFT Initialization}

Although our primary optimization is performed by end-to-end RL training, a lightweight SFT initialization is important for stabilizing training and eliciting consistent reasoning. 
In practice, a pretrained LLM may produce variable formats and shortcut rationales when facing long user histories and diverse item descriptions, which increases reward variance and degrades the signals consumed by the scoring head and the subsequent RL stage. 
We therefore warm-start the policy by imposing a structured self-reflective reasoning pattern, inspired by the empirical success of verification-type long-reasoning in recent reasoning-centric LLMs.

As shown in Figure~\ref{fig:selfcheck_template}, we introduce an instruction template that enforces a three-part output schema, consisting of an initial, thorough rationale, a self-check step that revisits the descriptive content provided in $c_u$ and identifies potential issues, and a final concise conclusion followed by a machine-readable decision. When facing different recommendation domains (e.g., books vs., movies), the template is shared across domains and only the placeholders (user context and candidate description) are instantiated with domain-specific fields (e.g., user profile, interaction history, item title, metadata, or textual content). 
Since standard recommendation datasets do not provide reasoning annotations, we use this template as a scaffold to \emph{synthesize} high-quality SFT supervision.

\noindent\textbf{SFT data construction.}
For each training instance, we instantiate the template with $(c_u, x_i)$ and query DeepSeek-R1 to generate a complete self-reflective response ending with a special decision token (\texttt{Recommend} / \texttt{Not Recommend}). We then validate DeepSeek-R1's decision against the ground-truth label of $x_i$ and keep only the \emph{correct} responses as our SFT data. %This yields instruction--response pairs that teach the backbone model to follow a stable long-form reasoning schema before RL.

We would like to highlight that the self-check section serves two purposes. First, it encourages the model to validate its own extraction of rationales from the recommendation context, which mitigates shallow reasoning under long contexts and improves decision consistency. Second, it yields more structured and information-dense hidden representations near the final decision token, benefiting our scoring head that consumes the content's hidden representation. This SFT stage is used only as initialization; the ranking objective is optimized in the subsequent RL stage.

\section{Experiments}

\subsection{Experimental Setup}

\subsubsection{Datasets and Metrics}
Following prior work on evaluation of LLM-based recommendation models \citep{you2025r2eclargerecommendermodels,bao2024decodingmattersaddressingamplification}, we evaluate on three classical Amazon datasets, including \textit{Musical Instruments}, \textit{Movies \& TV}, and \textit{Video Games}, together with a large-scale industrial dataset from a leading commercial advertising platform.
For all datasets, we treat observed user--item interactions as implicit positives and construct user sequences by sorting their interactions chronologically.
To match the LLM-based ranking setting and control maximum context length, we truncate each user history to the most recent $L{=}20$ interactions.
In Amazon datasets, we further restrict interactions to a fixed time window (Oct.\ 2021--Oct.\ 2023) to reduce temporal drift and alleviate item availability mismatch.

Table~\ref{tab:dataset_stats} reports dataset statistics after preprocessing.
Compared with the Amazon datasets, the industrial dataset is substantially larger in both item space and interaction volume, and exhibits a much more pronounced item cold-start regime (i.e., a large fraction of items appear rarely in training but are still encountered at test time), which makes user preference inference and robust recommendation much more challenging.

Across all datasets, we split users into train/validation/test sets with an 8:1:1 ratio.
We adopt a fixed candidate-set evaluation with $K{=}20$ candidates per instance (one ground-truth positive and 19 negatives), and report NDCG@$\{1,5,10\}$ over the induced ranked lists in all datasets. Unless stated otherwise, significance markers \sigone/\sigtwo/\sigthree correspond to $p<0.05/0.01/0.001$ under a two-sided paired Wilcoxon signed-rank test.

% \subsubsection{Task Setting}
% We formulate recommendation as a top-$K$ ranking problem over a fixed candidate set.
% For each evaluation instance, we construct a candidate pool of size $K{=}20$, consisting of exactly one ground-truth positive item and 19 negative items.
% The model outputs a complete ranking over the 20 candidates.
% This protocol is adopted for two reasons: (i) it matches the inference interface of LLM-based rankers that evaluate a candidate set under the same user context; and (ii) it enables consistent computation of listwise metrics and direct optimization toward the target metric in our RL setting.

% Unless otherwise specified, we evaluate on 3{,}000 sampled test instances per dataset for efficiency and consistent comparison across variants.
% We note that this setting can be challenging for prompted LLM baselines, which must produce a stable total order over non-trivial candidate lists under long contexts without task-specific post-training.

\subsubsection{Baselines}
To ensure a comprehensive evaluation, we compare \model with both conventional sequential recommenders and recent LLM-based recommenders.

\noindent\textbf{$\bullet$ Conventional recommenders.}
\textbf{SASRec}~\citep{kang2018selfattentivesequentialrecommendation} is a Transformer-based sequential recommender that models user behavior as an ordered interaction sequence and predicts the next item via self-attention.
\textbf{BERT4Rec}~\citep{sun2019bert4recsequentialrecommendationbidirectional} adopts bidirectional self-attention with masked-item prediction to capture contextual dependencies in user histories.
\textbf{LightGCN}~\citep{he2020lightgcnsimplifyingpoweringgraph} is a strong collaborative filtering model that performs simplified graph convolution over the user--item bipartite graph to learn interaction-driven embeddings.

\noindent\textbf{$\bullet$ LLM-based baselines.}
\textbf{DeepSeek-R1}~\citep{Guo_2025} is a general-purpose reasoning LLM; in our evaluation, we \emph{directly query} the model with the user history and the candidate set and let it output a ranked list, without any task-specific training.
\textbf{Prompt4NR}~\citep{Zhang_2023} formulates recommendation as a prompt-driven ranking/prediction problem, leveraging instruction-style prompting to elicit preference judgments from language models.
\textbf{TALLRec}~\citep{tallrec2023} proposes an efficient tuning paradigm to align LLMs with recommendation objectives, improving task alignment while controlling adaptation cost.
\textbf{R$^2$EC}~\citep{you2025r2eclargerecommendermodels} targets ``large recommender models with reasoning'' by integrating explicit reasoning traces with recommendation learning to better connect textual deliberation to ranking outcomes.

When official implementations are available, we use the released code and recommended hyper-parameters; otherwise, we conduct limited validation-set tuning to ensure stable performance. All baselines are evaluated using the same candidate-set protocol.

\begin{table}[t]
\centering
\caption{Statistics of datasets used in the experiments.}
\label{tab:dataset_stats}
% \resizebox{\linewidth}{!}
{
\begin{tabular}{lrrrr}
\toprule
\textbf{Dataset} & \textbf{\#Users} & \textbf{\#Items} & \textbf{\#Interactions} \\
\midrule
Musical Instruments & 13,031 & 10,326 & 49,171 \\
Movies \& TV         & 19,606 & 30,809 & 83,047 \\
Video Games          & 12,417 & 7,457  & 42,305 \\
Industrial           & 99,890 & 1,321,290 & 2,051,891 \\
\bottomrule
\end{tabular}}
\end{table}

\subsubsection{Backbone and Training Variants}
Our \model{} framework is built on Qwen-2.5-Instruct (1.5B and 3B) as the backbone LLM. To control for backbone capacity and isolate algorithmic differences, all trainable LLM-based baselines (Prompt4NR, TALLRec, and R$^2$EC) are implemented and trained on the same 3B backbone as ours, following their standard training recipes.
We report results with \textbf{SFT \& RL} by default, which warm-starts the model with SFT using the structured self-check format and then performs end-to-end RL post-training that optimizes NDCG@10 via our Plackett--Luce surrogate.
We also include an \textbf{RL-only} variant in the ablation study to isolate the effect of SFT initialization.
All ablation and in-depth analyses are conducted on the Qwen-2.5-3B-Instruct version of \model{} unless explicitly stated otherwise, ensuring consistent backbone capacity throughout the analysis.

\begin{table*}[t]
\centering
\caption{Comparison of recommendation performance on amazon and industrial datasets, where the performance is measured under different levels of NDCG. 
%\sigone/\sigtwo/\sigthree{} indicate statistically significant improvements over the second-best method on the corresponding dataset under a two-sided Wilcoxon signed-rank test.
}
\label{tab:main_results}
\resizebox{\textwidth}{!}
{
\begin{tabular}{lcccccccccccc}
\toprule
\multirow{2}{*}{\textbf{Method}} 
& \multicolumn{3}{c}{\textbf{Musical Instruments}}
& \multicolumn{3}{c}{\textbf{Movies \& TV}}
& \multicolumn{3}{c}{\textbf{Video Games}}
& \multicolumn{3}{c}{\textbf{Industrial}} \\
\cmidrule(lr){2-4} \cmidrule(lr){5-7} \cmidrule(lr){8-10} \cmidrule(lr){11-13}
& @1 & @5 & @10
& @1 & @5 & @10
& @1 & @5 & @10
& @1 & @5 & @10 \\
\midrule
\multicolumn{13}{l}{\textit{Traditional Recommendation Models}} \\
SASRec        & 0.257 & 0.416 & 0.468 & 0.107 & 0.205 & 0.258 & 0.223 & 0.332 & 0.366 & 0.072 & 0.115 & 0.201 \\                                                   BERT4Rec      & 0.218 & 0.328 & 0.422 & 0.065 & 0.148 & 0.228 & 0.166 & 0.267 & 0.304 & 0.095 & 0.168 & 0.229 \\                                                   LightGCN      & 0.248 & 0.402 & 0.455 & 0.102 & 0.198 & 0.247 & 0.215 & 0.320 & 0.356 & 0.051 & 0.143 & 0.238 \\ 
\midrule
\multicolumn{13}{l}{\textit{LLM-based}} \\
DeepSeek-R1 (671B)     
& 0.198 & 0.341 & 0.512 
& 0.182 & 0.476 & \textbf{0.619}\sigthree 
& 0.184 & 0.317 & 0.561 
& 0.292 & 0.438 & 0.500 \\
Prompt4NR    
& 0.173 & 0.305 & 0.421 
& 0.171 & 0.352 & 0.464 
& 0.161 & 0.289 & 0.432 
& 0.102 & 0.241 & 0.446 \\
TALLRec      
& \textbf{0.302}\sigone & 0.412 & 0.458 
& 0.156 & 0.268 & 0.354 
& 0.261 & 0.384 & 0.412 
& 0.126 & 0.281 & 0.402 \\
R$^2$EC      
& 0.289 & 0.408 & 0.449 
& 0.288 & 0.417 & 0.502 
& 0.241 & 0.358 & 0.468 
& 0.239 & 0.301 & 0.498 \\
\midrule
\multicolumn{13}{l}{\textit{Ours}} \\
\model (1.5B-Instruct)
& 0.234 & 0.453 & 0.514 
& 0.257 & 0.426 & 0.486 
& 0.279 & 0.487 & 0.549 
& 0.476 & 0.596 & 0.759 \\
\model (3B-Instruct)
& 0.279 & \textbf{0.495}\sigthree & \textbf{0.564}\sigthree 
& \textbf{0.338}\sigthree & \textbf{0.520}\sigthree & 0.581
& \textbf{0.330}\sigthree & \textbf{0.544}\sigthree & \textbf{0.596}\sigthree
& \textbf{0.524}\sigthree & \textbf{0.639}\sigthree & \textbf{0.818}\sigthree \\
\bottomrule
\end{tabular}
}
\end{table*}

\subsection{Main Results}

Table~\ref{tab:main_results} reports the overall performance on three Amazon datasets and the industrial dataset. We highlight four key findings.

\noindent\textbf{1) Overall effectiveness and listwise alignment.}
Across datasets, \model consistently achieves the best or near-best NDCG@10, demonstrating that (i) item-wise structured reasoning provide informative signals for relevance scoring, and (ii) directly optimizing listwise utility via our end-to-end RL training translates reasoning quality into ranking gains.

\noindent\textbf{2) Robustness on industrial data.}
While conventional recommenders remain competitive on some Amazon datasets, they degrade substantially on the industrial dataset, where exposure bias and noisier implicit negatives are much more pronounced.
In contrast, \model improves NDCG@10 from 0.500 (DeepSeek-R1) to 0.818, suggesting that grounding candidate assessment in user--item evidence and optimizing listwise ranking quality yields stronger robustness under real-world distribution shift.

\noindent\textbf{3) Domain dependence and world knowledge.}
On \textit{Movies \& TV} dataset, a larger backbone model (e.g., DeepSeek-R1) with stronger world knowledge can be advantageous, likely because accurate preference inference often requires richer entity-level associations and commonsense knowledge.
Despite using a compact 3B model, \model remains competitive on this domain, indicating that reasoning-to-rank training partially compensates for limited parametric knowledge.

\noindent\textbf{4) Metric focus.}
Since our post-training objective targets NDCG@10, \model{} prioritize improving the quality of the top-10 ranked list.
Accordingly, methods optimized for NDCG@10 may not always dominate when evaluating NDCG@1 or NDCG@5.

\begin{table}[t]
\centering
\caption{Joint training vs.\ scoring head-only training (Qwen2.5-3B-Instruct, RL-only) where the results are reported under NDCG@10.
%and \sigthree{} indicates p-value<0.001 under a two-sided paired Wilcoxon signed-rank test.
}
\label{tab:ablation_joint}
% \resizebox{\linewidth}{!}
{
\begin{tabular}{l l c c c c}
\toprule
\textbf{Variants}  & \textbf{Musical} & \textbf{Movies \& TV} & \textbf{Video Games} & \textbf{Industrial} \\
\midrule
MLP-only  & 0.368 & 0.322 & 0.397 & 0.528 \\
LLM+MLP   & \textbf{0.541}\sigthree  & \textbf{0.562}\sigthree  & \textbf{0.515}\sigthree  & \textbf{0.709}\sigthree \\
\bottomrule
\end{tabular}}
\end{table}

\subsection{Ablation Study}
We conduct ablation studies on Qwen2.5-3B-Instruct to investigate the contributions from the key components in \model{} design.
Unless otherwise specified, all variants are evaluated under the same 20-candidate protocol and compared on all datasets.
Since our training objective targets NDCG@10, we report NDCG@10 in the main ablation results for clarity; and we observe consistent result patterns under NDCG@1 and NDCG@5 metrics.

\subsubsection{Effect of Joint Training with the LLM}
We first examine whether the gains can be achieved by simply training the lightweight scoring head on top of a frozen LLM, versus end-to-end joint training of both the LLM backbone and the head under the same \textbf{RL-only} setting.
Table~\ref{tab:ablation_joint} shows that \textbf{MLP-only} (freezing the backbone LLM) achieves reasonable performance, suggesting that pretrained representations can already encode some level of preference-related signals.
However, jointly updating the \textbf{LLM+MLP} yields substantially larger improvements across all domains.
This indicates that the key driver is reasoning alignment: the backbone LLM must adapt its internal reasoning logic to better reflect final ranking utility, rather than relying on a shallow scorer over fixed perspectives.

\subsubsection{Effect of SFT Initialization}
Next, we study whether the self-reflective SFT initialization is necessary before RL.
We compare \textbf{RL-only} (no SFT initialization) with \textbf{SFT\ \&\ RL}. As shown in Table~\ref{tab:ablation_sft}, SFT initialization consistently improves NDCG@10 across all domains.
This suggests that enforcing a stable and informative reasoning schema prior to RL provides more consistent and informative decision signals for the scoring head, which complements listwise RL alignment.

\begin{table}[t]
\centering
\caption{Effect of SFT cold-start (Qwen2.5-3B-Instruct) where the results are reported under NDCG@10
%and \sigone/\sigtwo/\sigthree{} indicate corresponding levels of p-value under a two-sided paired Wilcoxon signed-rank test.
}
\label{tab:ablation_sft}
% \resizebox{\linewidth}{!}
{
\begin{tabular}{l l c c c c}
\toprule
\textbf{Variants} & \textbf{Musical} & \textbf{Movies \& TV} & \textbf{Video Games} & \textbf{Industrial} \\
\midrule
RL-only    & 0.541 & 0.562 & 0.515 & 0.709 \\
SFT \& RL  & \textbf{0.564}\sigone & \textbf{0.581}\sigtwo & \textbf{0.596}\sigthree  & \textbf{0.818}\sigthree  \\
\bottomrule
\end{tabular}}
\end{table}

\begin{figure*}[t]
\centering
\begin{subfigure}[t]{0.32\textwidth}
  \centering
  \includegraphics[width=\linewidth]{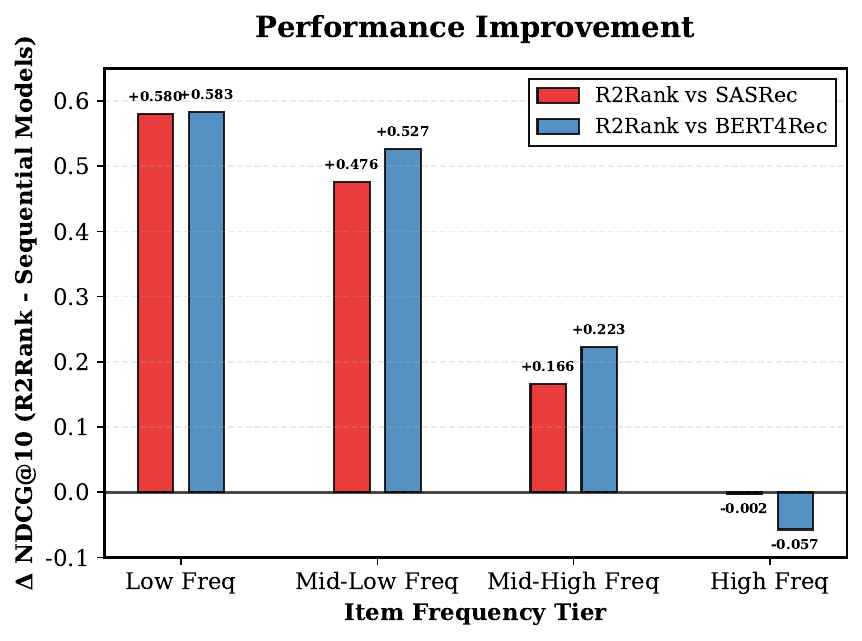}
  \caption{vs.\ sequential models (Video Games)}
  \label{fig:cold_start_a}
\end{subfigure}\hfill
\begin{subfigure}[t]{0.32\textwidth}
  \centering
  \includegraphics[width=\linewidth]{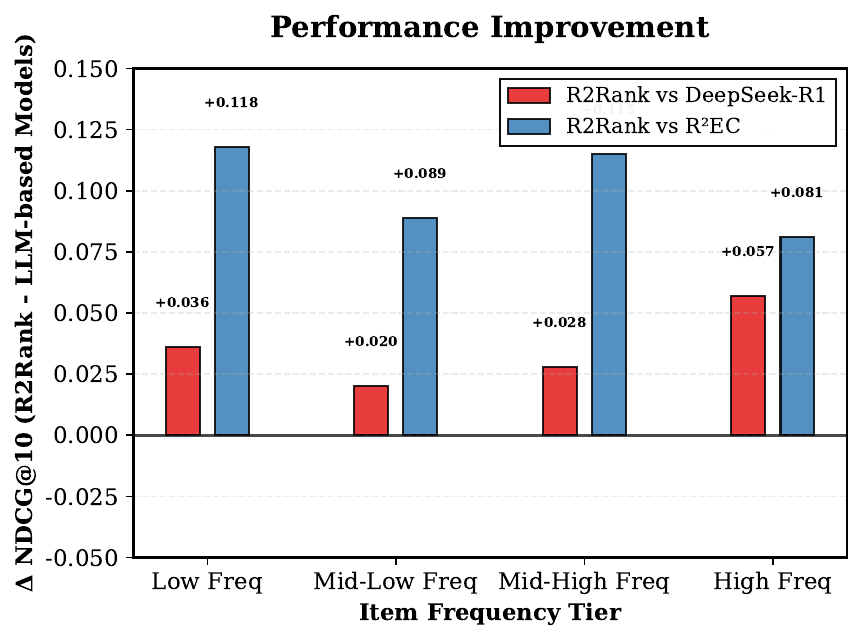}
  \caption{vs.\ LLM-based models (Video Games)}
  \label{fig:cold_start_b}
\end{subfigure}\hfill
% \begin{subfigure}
%   \centering
%   \includegraphics[width=\linewidth]{figures/item_cold_start_and_frequency_stratification2.pdf}
%   \caption{avg.\ frequency per tier}
%   \label{fig:cold_start_c}
% \end{subfigure}
\begin{subfigure}[t]{0.32\textwidth}
\centering
\includegraphics[width=\linewidth]
{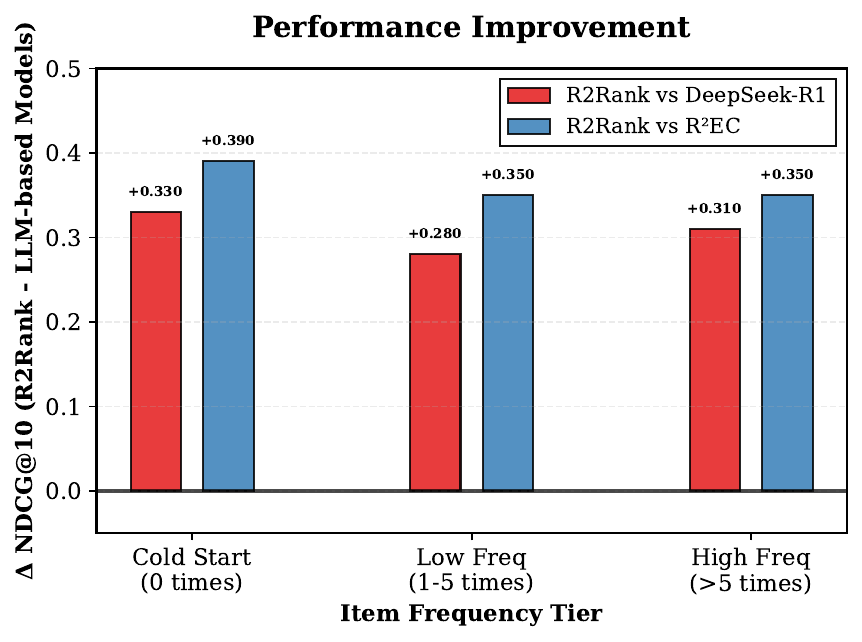}
\caption{vs.\ LLM-based models (Industrial)}
\label{fig:improvement_industrial}
\end{subfigure}

\caption{Comparison under item cold-start settings on \textit{Video Games} and \textit{Industrial} datasets.}
\label{fig:cold_start}
\vspace{-2mm}
\end{figure*}

\subsubsection{Effect of Explicit CoT Output}
Finally, we study whether explicitly generating chain-of-thought (CoT) during training contributes to ranking quality.
Here, \textbf{With CoT} corresponds to our default training setup (i.e., the standard prompt/format used throughout this paper), where the model is allowed to output a rationale followed by the final decision.
In contrast, \textbf{Without CoT} changes only the RL training prompt to force the model to output the final decision token(s) only, suppressing intermediate reasoning text, while keeping the training pipeline unchanged.
As shown in Table~\ref{tab:ablation_cot}, allowing CoT generation during RL training yields higher NDCG@10 across all datasets.
This suggests that exposing intermediate reasoning provides a richer and more structured training signal for aligning the backbone representations used by the scoring head, thereby improving the resulting listwise ranking utility.

\begin{table}[t]
\centering
\caption{Effect of explicit CoT output (Qwen2.5-3B-Instruct) where the results are reported by NDCG@10.
%and \sigone/\sigtwo/\sigthree{} indicate the corresponding levels of p-value under a two-sided paired Wilcoxon signed-rank test.
}
\label{tab:ablation_cot}
% \resizebox{\linewidth}{!}
{
\begin{tabular}{l l c c c c}
\toprule
\textbf{Variants} & \textbf{Musical} & \textbf{Movies \& TV} & \textbf{Video Games} & \textbf{Industrial} \\
\midrule
Without CoT  & 0.545 & 0.515 & 0.458 & 0.758 \\
With CoT    &  \textbf{0.564}\sigone  & \textbf{0.581}\sigthree  & \textbf{0.596}\sigthree  & \textbf{0.818}\sigthree  \\
\bottomrule
\end{tabular}}
\end{table}

\subsection{In-Depth Analyses}

\subsubsection{Cross-Dataset Generalization}
We evaluate cross-domain generalization by training \model on one Amazon dataset and directly evaluating the model on another dataset without any fine-tuning.
Table~\ref{tab:cross_domain} reports NDCG@10, where diagonal entries are in-domain performance and off-diagonal entries quantify cross-domain transfer.
Overall, \model exhibits encouraging cross-domain generalization: models trained on one domain retain non-trivial recommendation quality when evaluated on other domains (e.g., \textit{Video Games}$\rightarrow$\textit{Movies \& TV} and \textit{Movies \& TV}$\rightarrow$\textit{Video Games}).
These results suggest that the learned reasoning-to-rank policy captures transferable user--item compatibility signals beyond domain-specific co-occurrence patterns, consistent with our design of independent user--item judgments and listwise alignment.

\subsubsection{Item Cold-Start and Frequency Stratification}
To analyze models' robustness under item cold-start, we stratify test instances by their frequency of being the ground-truth positive item in the training set, and report the results obtained in the \textit{Video Games} dataset.
Specifically, we partition positive items into four equal-sized bins by their frequencies in the training set and evaluate recommendation quality within each tier.
Figure~\ref{fig:cold_start_a} reports the NDCG@10 improvement of \model over strong sequential baselines (SASRec and BERT4Rec), and in Figure~\ref{fig:cold_start_b}, we compare against two representative LLM-based recommenders (DeepSeek-R1 and R$^2$ec), across the four frequency tiers accordingly.

\begin{table}[t]
\centering
\caption{Cross-dataset generalization evaluated by NDCG@10. Models are trained on one dataset (rows) and directly evaluated on another (columns) without any other fine-tuning.}
\label{tab:cross_domain}
%\resizebox{\linewidth}{!}
{
\begin{tabular}{lccc}
\toprule
\textbf{Train $\backslash$ Test} & \textbf{Musical} & \textbf{Video Games} & \textbf{Movies \& TV} \\
\midrule
\textbf{Musical} & 0.564 & 0.491 & 0.412 \\
\textbf{Video Games} & 0.532 & 0.596 & 0.551 \\
\textbf{Movies \& TV} & 0.463 & 0.494 & 0.581 \\
\bottomrule
\end{tabular}}
\end{table}

Two observations stand out.
First, \model achieves the most pronounced gains on the lowest-frequency tiers, indicating strong robustness in cold-start regimes where the target item is rarely observed in training.
Notably, the advantage also holds against LLM-based baselines (Figure~\ref{fig:cold_start_b}), suggesting that the improvement is not a generic benefit of using an LLM, but is driven by our end-to-end training of the LLM.
Second, when the item frequency is very high, \model maintains a nearly comparable level, while the incremental gains over traditional baselines naturally shrink.
In this regime, conventional recommenders already benefit from abundant interaction signals and strong popularity priors, so the marginal advantage of explicit reasoning shrinks.
%Overall, the frequency-stratified results provide direct evidence that \model remains effective for rare items while staying competitive for frequent items, supporting its robustness to item cold-start.

Properly addressing cold-start is crucial in industrial settings, where the item space is huge: as reflected in our industrial dataset, \textbf{56.0\%} of test positives never appear in the training set.
As the conventional recommenders that rely heavily on historical co-occurrence signals are known to be ineffective in to handle cold-start, we focus on comparing \model against two LLM-based baselines mentioned above in our industrial dataset to understand the value of reasoning-driven recommendation under severe sparsity.

\begin{figure*}[t]
  \centering
  \begin{subfigure}[b]{0.49\textwidth}
    \centering
    \includegraphics[width=\linewidth]{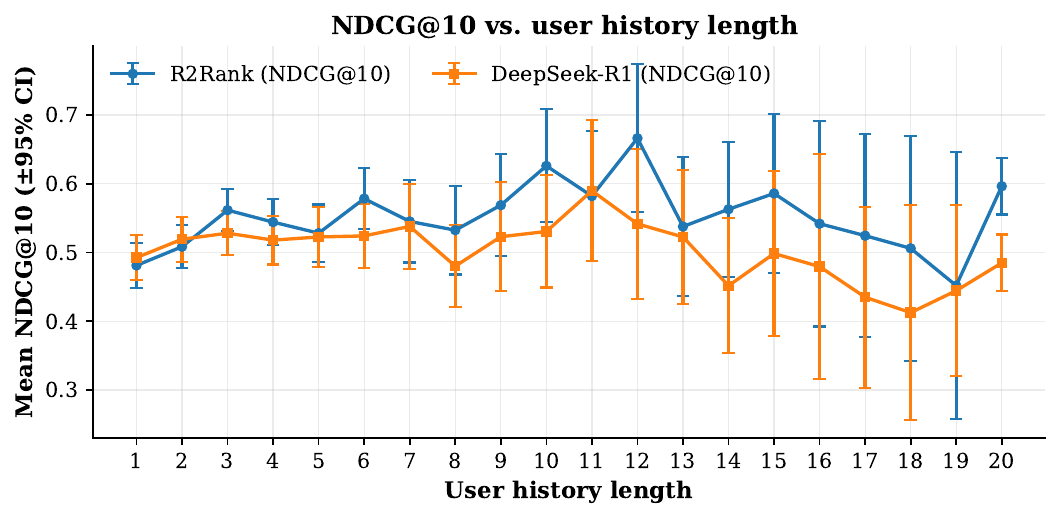}
    \caption{\text{Musical Instruments}}
    \label{fig:history_len_amazon}
  \end{subfigure}
  \hfill
  \begin{subfigure}[b]{0.49\textwidth}
    \centering
    \includegraphics[width=\linewidth]{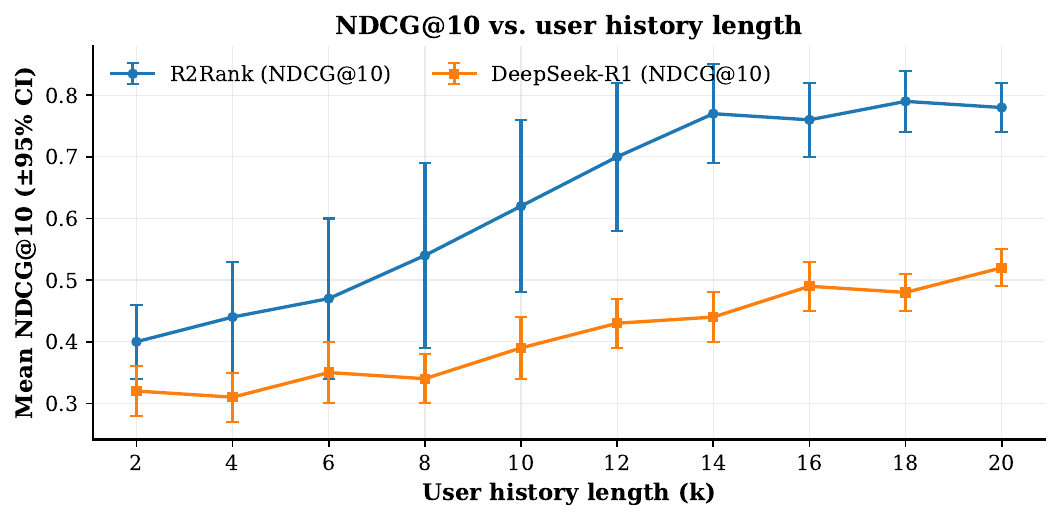}
    \caption{\text{Industrial}}
    \label{fig:history_len_industrial}
  \end{subfigure}
  \caption{Influence of user history length on recommendation quality.}
  \label{fig:history_length}
  \vspace{-2mm}
\end{figure*}

Instead of partitioning positives into equal-sized bins as in the Amazon dataset, we stratify industrial test instances by the training frequency of the ground-truth positive item into three practically motivated groups: cold-start (items never appeared in training; \textbf{56.0\%}), low frequency (occurrence between 1 and 5; \textbf{18.7\%}), and high frequency (all the rest items; \textbf{25.3\%}). The comparison against LLM-based models are summarized in Figure~\ref{fig:improvement_industrial}.

Figure~\ref{fig:improvement_industrial} suggests that the tier-wise gaps are generally consistent, likely because LLM-based baselines already provide strong semantic priors for rare items; and the end-to-end training in \model{} enables the model to best leverage the world knowledge and contextual information for recommendation. 
%This is consistent with our design goal: listwise post-training aligns the model’s pointwise judgments with ranking utility, which becomes most beneficial when the target item has no reliable co-occurrence evidence in training.

\subsubsection{Influence from User History Length}
We analyze the impact of user history length by stratifying test instances according to the number of historical interactions observed in the user and reporting mean NDCG@10 under each length category.
Figure~\ref{fig:history_length} shows the results (mean $\pm$ 95\% CI) on \textit{Musical Instruments} and \textit{Industrial} datasets.

Across different lengths of user history, \model maintains stable performance and competitiveness against (and often above) DeepSeek-R1 (the strongest baseline on this dataset).
This behavior is consistent with our design for the reasoning pattern: the self-reflecting reasoning encourages evidence-aware and consistent relevance assessment, producing more stable decision signals for the scoring head under long contexts.
Together with recommendation utility oriented RL training, these components help \model reliably extract and utilize relevant preference signals across a wide range of user history lengths.

% \begin{figure}[t]
% \centering
% \includegraphics[width=\linewidth]{figures/user_history_length_influence.pdf}
% \caption{NDCG@10 versus user history length on \textit{Musical Instruments} (mean $\pm$ 95\% CI).}
% \label{fig:history_len}
% \end{figure}

In the industrial dataset, long user histories are the norm: over \textbf{95\%} of users have more than 20 historical interactions.
To evaluate robustness under varying available context in this setting, we randomly sample \textbf{100} test instances whose truncated history length equals 20, and repeatedly construct inputs by keeping only the most recent $k$ interactions in user history.
We compare \model against DeepSeek-R1 under the same candidate-set evaluation protocol.

As shown in Figure~\ref{fig:history_len_industrial}, \model benefits more clearly from longer histories: its performance improves steadily as $k$ increases and remains consistently competitive with DeepSeek-R1 across all settings.
This result suggests that the proposed \model{} solution is better able to absorb and leverage incremental behavioral evidence in long-context regimes, which is especially important in industrial recommendation where rich user histories are ubiquitous.

% \begin{figure}[t]
% \centering
% \includegraphics[width=\linewidth]{figures/ndcg_vs_history_length.pdf}
% \caption{NDCG@10 versus available user history length on the \textit{Industrial} dataset. We sample 100 instances with truncated history length 20 and vary the retained suffix length $k \in \{2,4,\ldots,20\}$ (mean $\pm$ 95\% CI).}
% \label{fig:history_len_industrial}
% \end{figure}

\begin{figure}[t]
\centering
\includegraphics[width=0.7\linewidth]{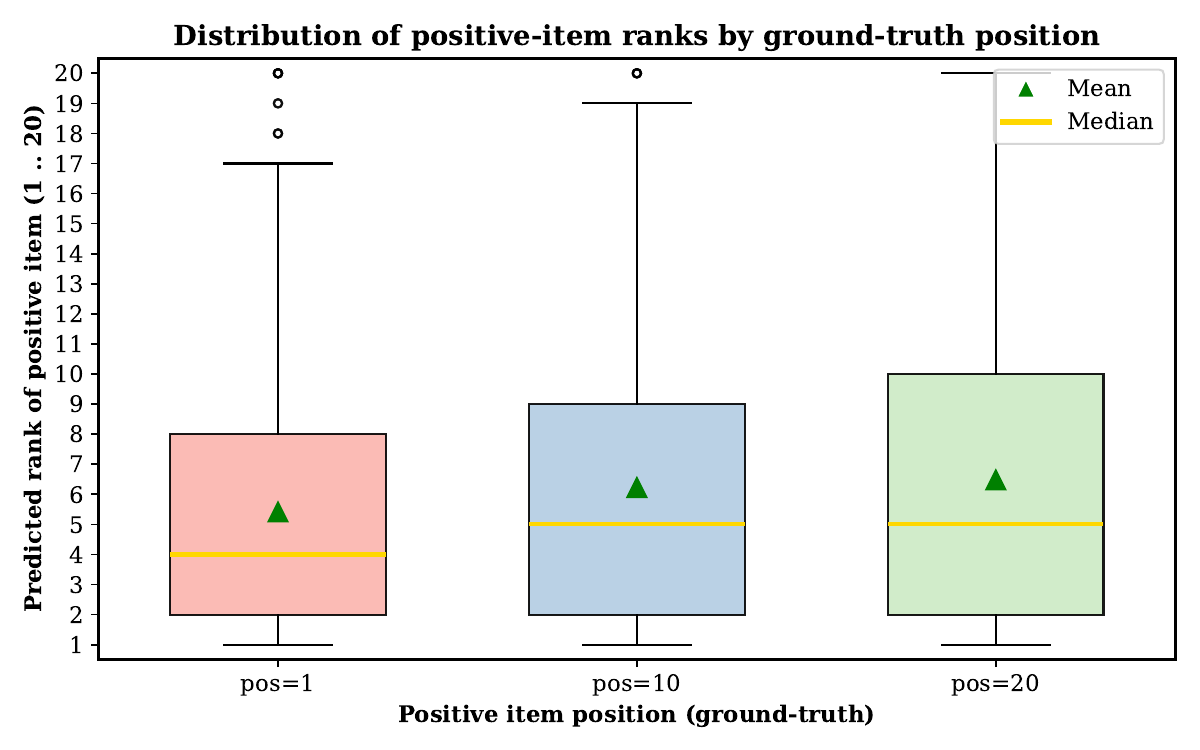}
\caption{Candidate position sensitivity of DeepSeek-R1 on the \textit{Musical Instruments} dataset.}
\label{fig:pos_sensitivity}
\end{figure}

\subsubsection{Case Study}
\begin{figure*}[t]
\centering
\includegraphics[width=0.955\textwidth]{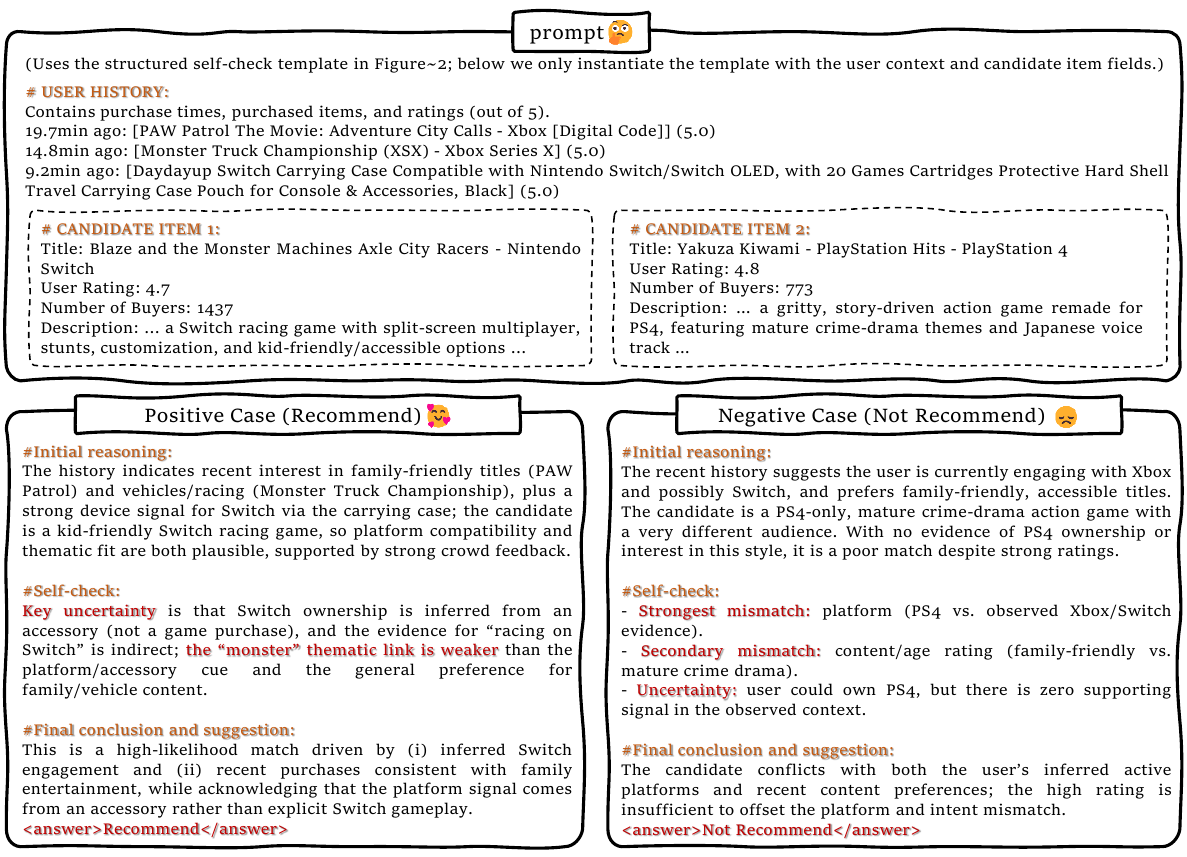}
\caption{Case study of \model on \textit{Video Games}: positive (Recommend) and negative (Not Recommend) reasoning examples produced under the structured self-reflective format. For brevity, we display condensed reasoning content while retaining the key evidence and self-check signals.}
\label{fig:case_study}
\vspace{-3mm}
\end{figure*}

To illustrate how \model improves the reasoning of user-item preference, we present a case study in Figure~\ref{fig:case_study}. Following the structured self-reflective template (Section~3.4), we show one positive and one negative item under the same user context with compacted rationales that retain the key evidence driving each decision. In the positive case, the model recognizes consistent game platform and intent cues (recent family-friendly purchases and Switch-related accessory), leading to a confident recommendation. In the negative case, the candidate is rejected primarily because of a strong platform mismatch (PS4-only) and a mismatch in tone/audience relative to the user’s observed preferences. These examples suggest that \model does not merely jump to the final answer, but encourages a more disciplined, evidence-first reasoning process that is easier to audit and more consistent with the resulting ranking behavior.

\subsection{Discussions}

\subsubsection{Candidate Position Sensitivity in Prompt-based Recommenders} 
A key motivation of \model is to mitigate the position bias in LLM input, where candidate ordering can inadvertently influence the model's output.
\model avoids this issue by performing independent user--item assessments: each candidate is evaluated in isolation under the same user context, and the final ranking is induced by sorting the scores produced by a shared scoring head.
As a result, the scoring function is permutation-invariant with respect to how candidates are presented.

To illustrate the practical impact of position bias in the baseline LLM solutions, we perform a controlled evaluation on \textit{Musical Instruments} using DeepSeek-R1.
We place the ground-truth positive item at different positions in the candidate list (position 1, 10, or 20) while keeping the remaining candidates unchanged, and evaluate the distribution of the predicted rank assigned to the positive item.
Figure~\ref{fig:pos_sensitivity} shows a clear positional impact: placing the positive item at the first position leads to noticeably better outcomes than placing it at later positions.
%, whereas the difference between positions 10 and 20 is comparatively smaller.
This suggests that simply feeding user history into an LLM can conflate relevance with presentation order, reinforcing our design choice of decoupling candidate evaluation from list presentation in \model.

\begin{table}[t]
\centering
\caption{Effect of history order randomization on \textit{Video Games} (\model, Qwen2.5-3B-Instruct, NDCG@10). ``Original'' uses the chronological history.}
%; ``Shuffled'' removes timestamps and permutes the history order.
\label{tab:history_shuffle}
%\resizebox{\linewidth}{!}
{
\begin{tabular}{lccccc}
\toprule
\textbf{Method} & \textbf{Avg} & \textbf{Std} & \textbf{Range} & \textbf{Original Avg} \\
\midrule
\model & 0.596 & 0.085 & 0.211 & 0.598 \\
SASRec & 0.296 & 0.030 & 0.062 & 0.297 \\
BERT4Rec & 0.229 & 0.017 & 0.036 & 0.230 \\
\bottomrule
\end{tabular}}
\end{table}

\subsubsection{Sensitivity to User History Order}
Modeling the sequential pattern in user's interaction history is known to be important for quality recommendations \citep{kang2018selfattentivesequentialrecommendation,sun2019bert4recsequentialrecommendationbidirectional}. But the influence from the ordering of items in user history is compounded with the position bias in LLMs, as the history is still serialized as a token sequence, and reordering it changes the conditioning context of the LLM, which can alter the generated rationale and the resulting score. 

%\model mitigates candidate-list position bias by evaluating candidates independently and inducing the final ranking via score sorting. However, this permutation invariance does not automatically extend to the user history: the history is still serialized as a token sequence, and reordering it changes the conditioning context of the LLM, which can alter the generated rationale and the resulting score. Therefore, while \model breaks position bias on the candidate side, it can still be affected by history-order (prompt) bias on the context side.

We probe this impact on the \textit{Video Games} dataset by randomly permuting the interaction history in the testing instances and comparing against the original chronological order. As shown in Table~\ref{tab:history_shuffle}, although the mean NDCG@10 is close across settings for \model as well as SASRec/BERT4Rec, the variance of \model{} performance is much larger than the baselines'. 
%This result should not be interpreted as implying that temporal order is unimportant; rather, under our current evaluation protocol (the truncated $L{=}20$ recent-history context), a substantial fraction of the predictive signal may already be captured by the content and composition of recent items, making multiple orderings yield similar average ranking quality.
%At the same time, \model exhibits non-trivial dispersion across permutations (Std/Range), indicating that rankings can still change materially when the same evidence is presented in a different order, which is consistent with token-level conditioning in prompt-based reasoning. 
Unfortunately, when examining the detailed results, the input with chronological order does not necessarily led to the best results in \model. 
Ideally, sensitivity to history order should reflect meaningful temporal effects (e.g., preference drift and recency), not artifacts of how the context is serialized. This motivates future work on reducing context presentation bias, such as order-robust history representations or training-time augmentations that encourage stability across alternative history serializations while preserving genuine temporal signals.

%\subsubsection{Limits of Independent User--Item Assessment}
%While independent user--item assessment is effective for reducing position bias and stabilizing reasoning under long contexts, it also introduces a subtle limitation: the model evaluates each candidate with its own locally grounded rationale, which may not always provide a globally consistent comparison standard across items.
%In realistic recommendation scenarios, different items can invite different evaluation foci (e.g., functionality, style, brand affinity, price sensitivity), and purely item-wise rationales may emphasize heterogeneous criteria.
%As a result, even when each individual judgment is plausible, the induced scores may reflect partially incomparable rationales across candidates, potentially affecting calibration and fine-grained ranking decisions.

%Addressing this issue likely requires a more principled notion of comparability across candidates, so that scores are grounded in consistent criteria.
%We consider this an open direction for future work in reasoning-to-rank recommenders.

\section{Conclusion}

%In this work, we study how to turn LLM reasoning into a trainable ranking decision rather than a post-hoc explanation. We propose Reasoning2Rank, an end-to-end framework that decouples recommendation into independent user--item reasoning judgments and a lightweight scoring head, and then aligns these reasoning-driven scores with listwise ranking utility via PPO-based learning-to-rank reinforcement learning under a Plackett--Luce surrogate. A self-check SFT warm start further stabilizes long-context reasoning and improves the quality of decision representations for ranking.

%Extensive experiments on three Amazon domains and a large-scale industrial dataset demonstrate consistent improvements over strong traditional and LLM-based baselines, with statistically significant gains on key benchmarks. Ablation and in-depth studies validate the necessity of joint LLM training, explicit CoT, and SFT initialization, and show that the resulting policy generalizes across domains and remains robust under item cold-start and varying user-history lengths. Our discussion further highlights practical implications and limitations, pointing to future work on mitigating history-order sensitivity in prompt-conditioned reasoning and on improving global comparability across candidates, while retaining permutation-invariant reasoning-to-rank behavior.

We presented Reasoning to Rank (\model), a novel end-to-end framework for LLM-based recommendation. By internalizing the ranking objective directly into the language model's generative process, we bridged the long-lasting gap between semantic reasoning and recommendation utility. 
Crucially, we established a reinforcement learning mechanism utilizing a Plackett–Luce differentiable surrogate, which enables rank-level credit assignment; this allows non-differentiable listwise metrics (e.g., NDCG) to effectively propagate back to the token-level reasoning process. And our approach leverages a self-reflective supervised fine-tuning strategy to initialize the model with robust, intent-aware inference capabilities. 
Extensive experiments across public benchmarks and a large-scale industrial dataset demonstrate \model consistently outperforms competitive baselines, validating the effectiveness of our design.

Our study also opens several meaningful new direction for further exploration. While independent user-item assessment is effective for reducing position bias, it also introduces a subtle limitation: the model evaluates each candidate with its own locally grounded rationale, which may not always provide a globally consistent comparison standard across items.
%In realistic recommendation scenarios, different items can invite different evaluation foci (e.g., functionality, style, brand affinity, price sensitivity), and purely item-wise rationales may emphasize heterogeneous criteria.
%As a result, even when each individual judgment is plausible, the induced scores may reflect partially incomparable rationales across candidates, potentially affecting calibration and fine-grained ranking decisions.
Addressing this issue likely requires a more principled notion of comparability across the reasoning process, so that scores are grounded in consistent criteria.
Besides, our experiment results suggest context modeling is compounded with item ordering in user history, which calls for finer grain design of LLM input. Layered position embeddings in the Transformer architecture could be the solution. 
Last but not least, exposing the generated rationales to users provides a viable way to foster trust and transparency. This would enable interactive feedback loops, allowing users to critique or correct the model's logic directly via natural language.
%We consider this an open direction for future work in reasoning-to-rank recommenders.
%and offering a scalable paradigm for the future development of reasoning-centric recommender systems.

\bibliographystyle{tmlr}
\bibliography{sample-base}

% \appendix
% \section{Appendix}

\end{document}